\documentclass[aps,pra,superscriptaddress,floatfix,twocolumn,longbibliography]{revtex4-2}
\usepackage[T1]{fontenc}
\usepackage[utf8]{inputenc}
\usepackage{float}
\usepackage{amsmath}
\usepackage{amsfonts}
\usepackage{graphicx}
\usepackage{xcolor}
\usepackage[english]{babel}
\usepackage{makecell}

\makeatletter

\newcommand{\pd}[1]{\hat{\psi}^\dagger_{#1}}
\newcommand{\pc}[1]{\hat{\psi}_{#1}}
\newcommand{\pds}[1]{\hat{\psi}^{\dagger 2}_{#1}}
\newcommand{\pcs}[1]{\hat{\psi}^{2}_{#1}}

\makeatother

\begin{document}
\title{Chiral and pair superfluidity in triangular ladder produced by state-dependent Kronig-Penney lattice} 

\author{Domantas Burba}
\email{domantas.burba@ff.vu.lt}
\affiliation{Institute of Theoretical Physics and Astronomy, Faculty of Physics, Vilnius University,  Saul\.{e}tekio 3, LT-10257 Vilnius, Lithuania.}
\author{Giedrius \v{Z}labys}
\affiliation{Quantum Systems Unit, Okinawa Institute of Science and Technology Graduate University, 904-0495 Okinawa, Japan}
\author{Dzmitry Viarbitski}
\affiliation{Institute of Theoretical Physics and Astronomy, Faculty of Physics, Vilnius University,  Saul\.{e}tekio 3, LT-10257 Vilnius, Lithuania.}
\author{Thomas Busch}
\affiliation{Quantum Systems Unit, Okinawa Institute of Science and Technology Graduate University, 904-0495 Okinawa, Japan}
\author{Gediminas Juzeli\={u}nas}
\affiliation{Institute of Theoretical Physics and Astronomy, Faculty of Physics, Vilnius University,  Saul\.{e}tekio 3, LT-10257 Vilnius, Lithuania.}

\date{\today}

\begin{abstract}
We propose a concrete realization of a triangular ladder for ultracold atoms, which simultaneously hosts geometric frustration and unusual two-body interactions, and in particular controllable pair hopping and density-induced tunneling. This is done by means of a spin-dependent Kronig-Penney lattice created using a spatially-dependent tripod-type atom-light coupling. We apply density matrix renormalization group (DMRG) calculations to derive the quantum phase diagram. We find that pair tunneling stabilizes a robust pair superfluid, characterized by power-law decay of pair correlations. Additionally, a chiral superfluid arises from frustration induced by competing nearest neighbor (NN) and next-nearest neighbor (NNN) tunnelings. Finally, in the high barrier regime, we map our system onto the XXZ spin model and find the exact phase transition points.
\end{abstract}

\maketitle

\section{Introduction}

Ultracold atoms in optical lattices~\cite{JakschZollerPRL1998, Jaksch2005Annals} currently provide one of the most clean and controllable platforms for quantum simulation of many-body systems~\cite{Lewenstein07, Bloch2008RMP}. They offer tunable interactions, flexible lattice geometries, and precise state preparation and detection. However, correlated hopping is negligible and geometric frustration is not present in conventional optical lattices. In this context, the use of state-dependent optical lattices, including the one considered in this work, introduces an additional and highly versatile method of control \cite{Bloch03PRL,Jaksch_2003,Gerbier_2010,Soltan-Panahi2011,Dalibard2011Nov,Anderson2020,Anisimovas2016,EdvinasScipost,Kubala2021,Burba2025CommPhys}. 

Geometric frustration and correlated tunneling processes are two powerful mechanisms for generating unconventional quantum phases in strongly interacting systems. Frustration can induce chiral order~\cite{Kolezhuk2008PRB}, macroscopic degeneracies~\cite{moessner2006geometrical}, and exotic magnetic states, such as spin liquids~\cite{Norman2020Science}. On the other hand, correlated hopping~\cite{Huber2013PRB, EckholtNJP2009, Zhou2009PRA,Kremer:17} (pair or density-induced tunneling) promotes coherent motion of pairs of particles instead of individual ones, which is relevant to graphene~\cite{Hamilton2013PRL},
double quantum wells~\cite{Hubert2019PRX},
and superconductors~\cite{Berg2009}. Although each mechanism has been extensively explored separately~\cite{EckholtPRA2008, EckholtNJP2009, LucaScipost2024, Huber2013PRB, Santos2025arXivChiral, Burba2025CommPhys}, physical platforms that simultaneously realize strong geometric frustration and sizable correlated hopping remain scarce.

In this work, we consider many-body properties of a state-dependent Kronig-Penney lattice with nearest-neighbor (NN) and next-nearest-neighbor (NNN) tunneling, where the NNN amplitude dominates and carries an opposite sign~\footnote{The single particle properties of the spin-dependent Kronig-Penney lattices created using the tripod scheme were considered in refs.~\cite{EdvinasScipost,Kubala2021}}. The resulting effective model is a frustrated triangular ladder with an effective $\pi$-flux per plaquette. Importantly, state-dependent interactions naturally generate sizable correlated tunneling processes, including pair hopping and density-induced tunneling, whose strengths are controlled by interaction asymmetries and overlaps of Wannier functions.

Using large-scale density matrix renormalization group (DMRG) simulations, we determine the many-body phase diagram of the corresponding extended Bose-Hubbard model. We identify Mott insulating (MI), density wave (DW), pair superfluid (PSF), and chiral superfluid (CSF) phases. Pair tunneling stabilizes a robust PSF characterized by quasi-long-range pair correlations and suppresses single-particle correlations, whereas geometric frustration induces spontaneous time-reversal symmetry breaking and the emergence of a CSF with long-range current correlations. In the high barrier regime, where single-particle tunneling is negligible, the system maps onto an antiferromagnetic XXZ spin-1/2 chain, allowing analytical determination of phase boundaries and providing a benchmark for the numerical results.

The manuscript is organized as follows. In Sec.~II, we introduce the tripod atom-light coupling scheme and derive the effective lattice Hamiltonian. In Sec.~III, we define the relevant order parameters and present the DMRG results for the ground-state phase diagram. Section IV discusses the mapping to the XXZ spin model in the high barrier regime and compares analytical predictions with numerical findings. Finally, Sec. V summarizes the results and outlines possible future directions.

\section{Formulation}
\label{sec:formulation}

\subsection{Hamiltonian}

\begin{figure}[h]
\includegraphics[width=1.02\columnwidth]{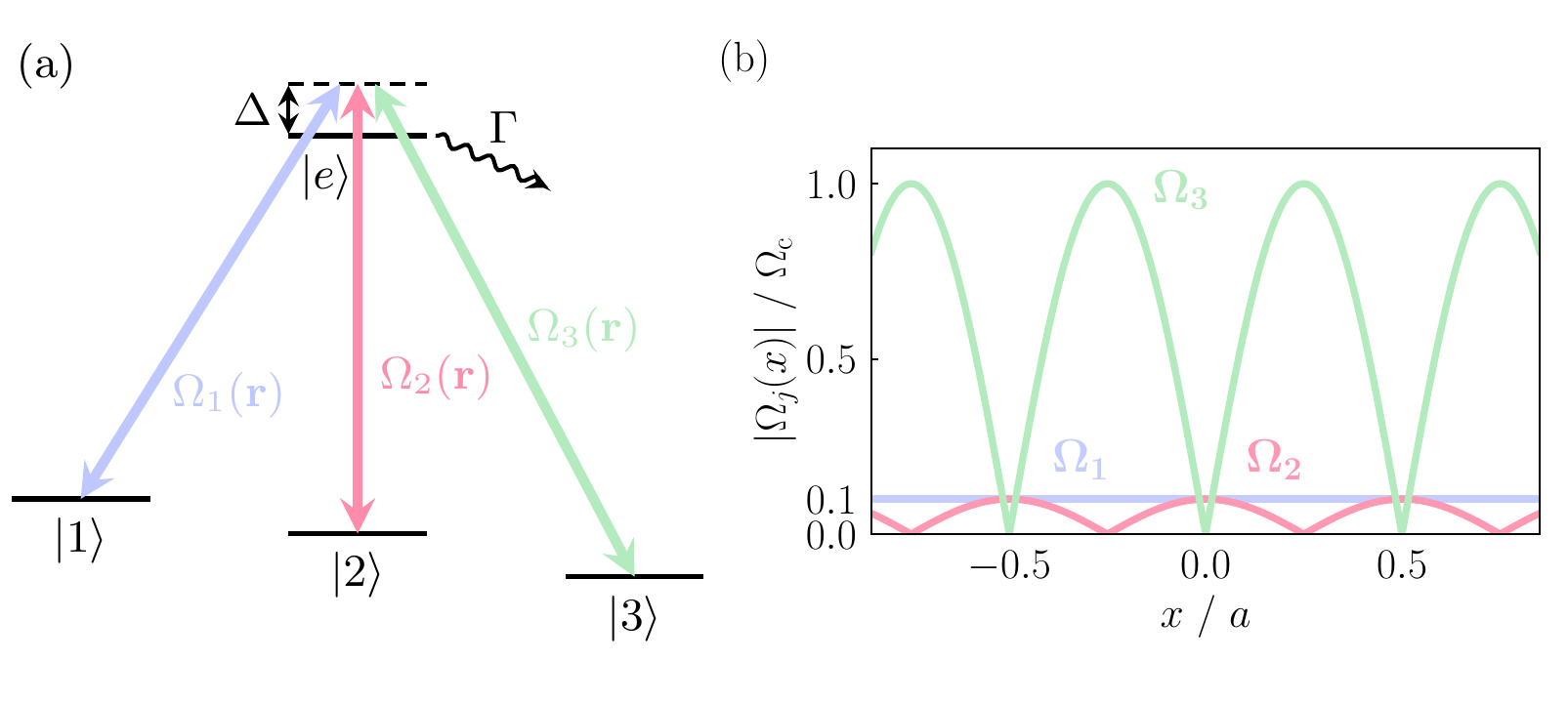}
\caption{\label{fig:FigTripod} (a) Tripod atom-light coupling configuration for ultracold atoms. 
(b) Spatial dependence of the  modulus of the Rabi frequencies $\Omega_{l}$ ($l\in \left\{1,2,3\right\}$) obtained from Eq.~\eqref{eq:Tripod-Omegas} for $\epsilon=0.1$.}
\end{figure}

We investigate a one-dimensional state-dependent optical lattice composed of sub-wavelength barriers that scatter atoms differently based on their internal states. Specifically, the internal atomic states are taken to be the two dark states realized by a tripod atom light coupling scheme \cite{Unanyan1998Oct,Ruseckas2005Jun,Dalibard2011Nov} which can create narrow state-dependent potential barriers \cite{EdvinasScipost,Kubala2021}. Such barrier lattices can be approximated as Kronig-Penney lattices~\cite{WenBin2025, Reshodko_2019, zlabys2025emergenttopologicalpropertiesspatially}. The full atomic Hamiltonian includes the kinetic energy term and the operator of the atom-light coupling in the tripod scheme
\begin{equation}
\hat{H}_{\mathrm{AL}} = \frac{\hat{p}^2}{2m}  +\hat{V}_{\mathrm{AL}},
\end{equation}
where the latter operator $\hat{V}_{\mathrm{AL}}$ 
reads in the rotating frame
\begin{equation}
\hat{V}_{\mathrm{AL}} = - \left(\Delta +\mathrm{i}\frac{\Gamma}{2} \right)|e \rangle \langle e|+\frac{1}{2}\sum_{l=1}^3 \left[\Omega_{l}(x)|0\rangle \langle l| + \mathrm{h.c.}\right].
\end{equation}
Here, each atomic ground-state $| l \rangle$ (with $l = 1,2,3$)
is coupled with the Rabi frequency $\Omega_j (x)$ to the same excited state $| e \rangle$ which is detuned by $\Delta$ and has a decay rate of $\Gamma$, as illustrated in Fig.~\ref{fig:FigTripod}(a).

Atoms coupled to light in the tripod scheme have two linearly independent dark states, representing superpositions of ground states that are not affected by radiation. The dark states are eigenstates of the atom-light coupling operator $\hat{V}_{\mathrm{AL}}$ with zero eigenenergy.  A frequent choice of an orthogonal pair of dark states is \cite{Unanyan1998Oct,Ruseckas2005Jun,Dalibard2011Nov}
\begin{equation}
\label{eq:DSbasis}
\begin{aligned}
|D_1 \rangle &= \sin \phi |1\rangle - \cos \phi |2 \rangle, \\
|D_2 \rangle &= \cos\theta \left(\cos \phi |1\rangle + \sin \phi |2 \rangle \right) - \sin \theta | 3 \rangle,
\end{aligned}
\end{equation}
where we have introduced the spherical coordinates for the Rabi frequency vector $\mathbf{\Omega}(x) = (\Omega_1(x),\Omega_2(x),\Omega_3(x))$ through $\cos \theta = \Omega_3(x)/\Omega(x)$ and $\tan \phi = \Omega_2(x)/\Omega_1(x)$, where $\Omega(x)\equiv|\mathbf{\Omega}(x)|$ is the total Rabi frequency. If the characteristic atomic center of mass energy (in frequency units) is small compared to the total Rabi frequency $\Omega(x)$, the adiabatic approximation can be used to restrict the system's dynamics to the manifold spanned by the dark states.

\begin{figure}[h]
\includegraphics[width=0.9\columnwidth]{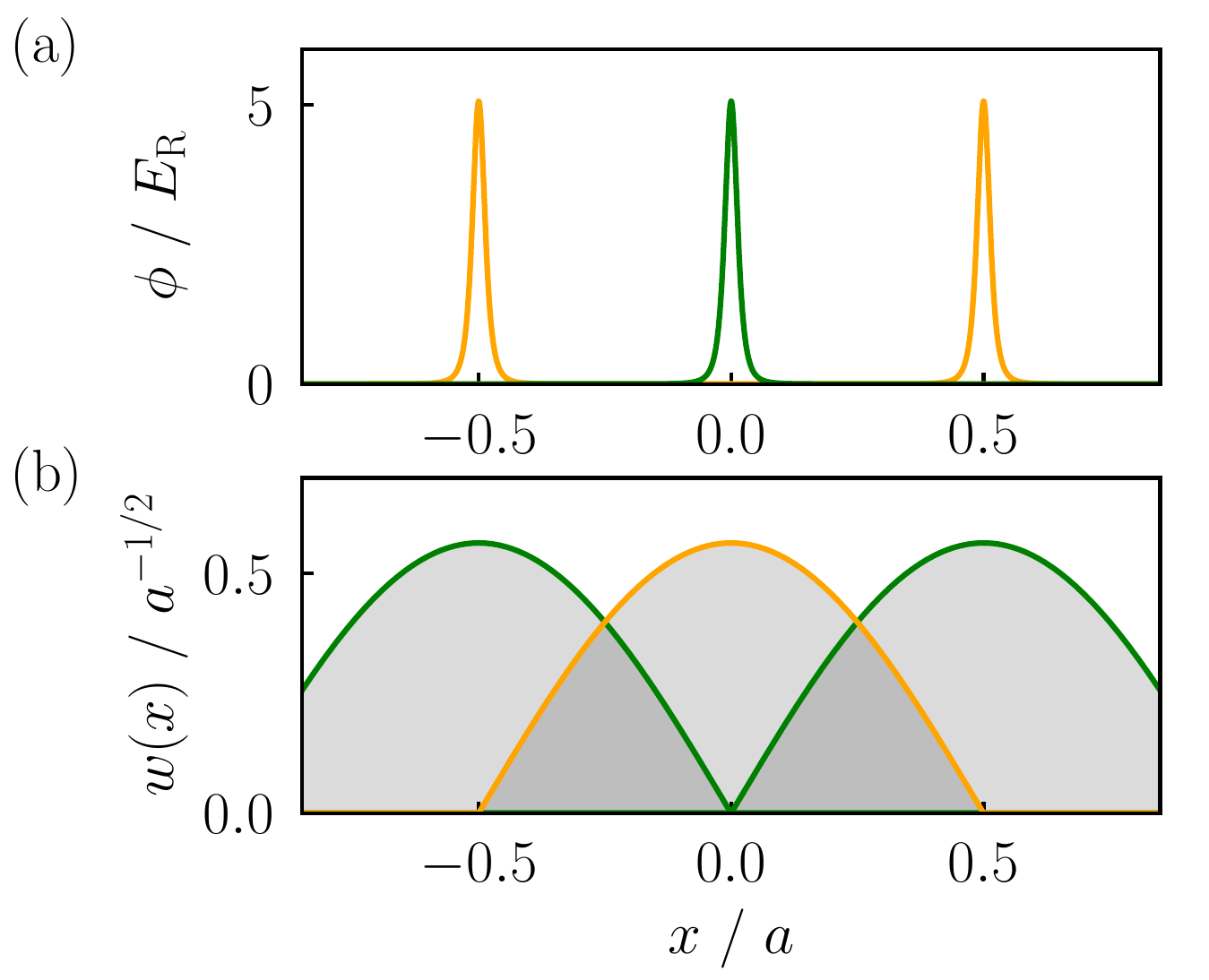}
\caption{\label{fig:FigBrickWall} (a) The geometric scalar potential $\phi\approx\hbar^2\theta^{\prime 2}/2m$  {\cite{EdvinasScipost}} 
in the units of the recoil energy $E_{\mathrm{R}}=\hbar^2k^2/2m$ for $\epsilon=\Omega_\mathrm{p}/\Omega_\mathrm{c}=0.1$. The potential represents a periodic array of barriers that act in the alternating manner on the symmetric (green) or antisymmetric (yellow) superposition of atomic ground states $|\pm\rangle$.   
(b) Schematic representation of the corresponding Wannier functions of the lowest (the $s$) Bloch band for atoms in the symmetric (green) or antisymmetric (yellow) superposition states $|\pm\rangle$.
}
\end{figure}

A periodic array of state-dependent subwavelength barriers can be created using the Rabi frequencies \cite{EdvinasScipost,Kubala2021} 
\begin{equation}
\label{eq:Tripod-Omegas}
\begin{aligned}
\Omega_1(x) &= \Omega_\mathrm{p}, \\
\Omega_2(x) &= \Omega_\mathrm{p} \cos(2\pi x/a), \\
\Omega_3(x) &= \Omega_\mathrm{c} \sin(2\pi x/a),
\end{aligned}
\end{equation}
where the amplitudes of $\Omega_{1,2}(x)$ should be much smaller than of the third Rabi frequency, $\epsilon=\Omega_\mathrm{p}/\Omega_\mathrm{c} \ll 1$, as illustrated in Fig.~\ref{fig:FigTripod}(b). Here, $a=2\pi/k$ is the optical wavelength and $k$ is the wavenumber.

The subwavelength barriers of the width 
$ \epsilon a$ 
are formed at the zero points of the Rabi frequency $\Omega_3(x)$ \cite{EdvinasScipost,Kubala2021} at $x=na/2$, where $n$ is an integer. The Kronig-Penney type barriers act on atoms in the dark state $|D_2 \rangle$, because this dark state experiences sudden changes around the zero points of $\Omega_3(x)$.
The dark state $|D_2 \rangle$ reduces to the symmetric and antisymmetric superposition of the atomic ground states, $|\pm\rangle=\left(|1\rangle\pm|2\rangle\right)/\sqrt{2}$, at the barriers corresponding to even and odd $n$, respectively. 
Thus, atoms in the  states $|\pm\rangle$  are scattered at every second barrier,
as shown in Fig.~\ref{fig:FigBrickWall}(a). Consequently, the atomic Wannier function $w_n(x)=w(x-na/2)$ centered at $x=na/2$ is localized between the neighboring barriers at $x=(n-1)a/2$ and $x=(n+1)a/2$. 
Such Wannier functions have been considered in detail in \cite{EdvinasScipost}.
We will concentrate on the situation where the atoms populate the lowest Bloch band and are described by the corresponding Wannier functions, depicted in Fig.~\ref{fig:FigBrickWall}(b).

\begin{figure}[h]
\includegraphics[width=1.02\columnwidth]{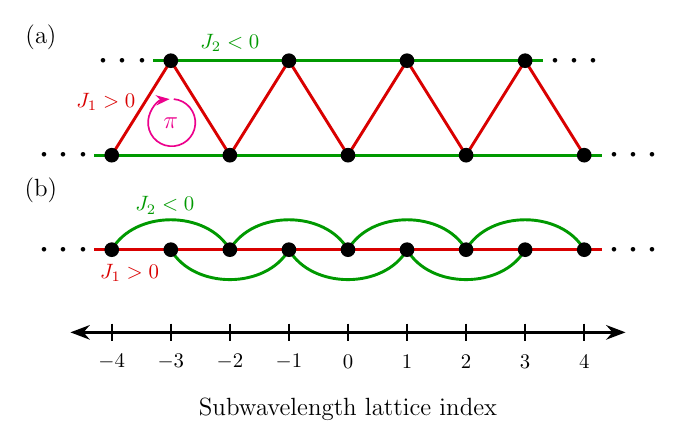}
\caption{\label{fig:FigTriangular} 
(a) Triangular ladder with opposite-sign single-particle hoppings $J_1$ and $J_2$, yielding effective geometric frustration. Consequently, each triangular plaquette has a $\pi$-flux. (b) The corresponding linearized 1D chain described by Hamiltonian of Eq.~\eqref{eq:WannierBasis} containing the NN and NNN couplings $J_1$ and $J_2$.
}
\end{figure}

The bosonic annihilation field operators $\hat{\psi}_\pm (x)$ corresponding to the atomic states $|\pm\rangle$ can be expanded in the Wannier function basis $w_n(x)=w(x-na/2)$, where $n \in 2\mathbb{Z}+1$ for the $|+\rangle$ state and $n \in 2\mathbb{Z}$ for the $|-\rangle$ state, i.e.
\begin{equation}
\label{eq:WannierBasis}
\begin{aligned}
\hat{\psi}_+(x) &= \sum_{j} w_{2j+1}(x) \hat{a}_{2j+1}, \\
\hat{\psi}_-(x) &= \sum_{j} w_{2j}(x) \hat{a}_{2j}.
\end{aligned}
\end{equation}
Here $\hat{a}_n$ is the operator for the annihilation of an atom characterized by the Wannier centered at $na/2$.

In the Wannier function basis for the lowest Bloch band, the present atomic system, including the atom-atom interaction, can be described by a Bose-Hubbard type Hamiltonian 
\begin{equation}
\label{eq:HMain}
\hat{H}=\hat{H}_{{\rm tun}} + \hat{H}_{{\rm int}}.
\end{equation}
The first term $\hat{H}_{{
\rm tun}}$ is the single particle (tunneling) Hamiltonian  described by the nearest-neighbor (NN) and next-nearest-neighbor (NNN) coupling with the strengths $J_1$ and $J_2$, respectively.
\begin{equation}
\label{eq:HTunMain}
\hat{H}_{\rm tun} =
     -J_{1} \sum_{j} \left( \hat{a}_{j}^{\dagger} \hat{a}_{j+1} + \text{h.c.} \right)
     -J_{2} \sum_{j} \left( \hat{a}_{j}^{\dagger} \hat{a}_{j+2} + \text{h.c.} \right).
\end{equation}
This one-dimensional (1D) lattice with NN and NNN tunnelings can be represented in terms of a triangular ladder shown in Fig.~\ref{fig:FigTriangular}(a). 

In previous work on the tripod scheme ~\cite{EdvinasScipost}, it was shown that the tunneling strengths are staggered for  the lowest Bloch band, i.e., $J_1>0$ and $J_2<0$. Notably, the NNN coupling is stronger than NN, i.e., $|J_2|>|J_1|$, which leads to significant geometric frustration. 
This is because of specific features of the current lattice illustrated in Fig.~\ref{fig:FigBrickWall}, where the NN coupling is due to the overlap between the neighboring Wannier functions corresponding to different internal states $|\pm\rangle$ and $|\mp\rangle$, so the NN tunneling is non-zero only because of a tiny admixture between these states. On the other hand, the NNN coupling is due to the tunneling over the barriers without the change in the internal state.
Note that one can control the tunneling amplitudes by changing the barrier heights via the ratio of Rabi frequency amplitudes $\Omega_{\mathrm{p}}/\Omega_{\mathrm{c}}$~\cite{Jendrzejewski2016,Zoller2016,EdvinasScipost,Kubala2021}.
The single-particle problem of state-dependent tripod barriers has been explicitly treated in Ref.~\cite{EdvinasScipost,Kubala2021}.

The second term in the Hamiltonian (\ref{eq:HMain}) describes the atom-atom interaction and is given by (see Appendix \ref{appendix:HIntDerivation} for the derivation details)
\begin{equation}
\label{eq:HIntMain}
\begin{aligned}
\hat{H}_{\rm int} &=
	\frac{U}{2} \sum_{j} \hat{n}_{j} (\hat{n}_{j} - 1) 
    + V \sum_{j} \hat{n}_{j} \hat{n}_{j+1} \\
    & - D \sum_{j} \hat{a}_{j}^{\dagger} \hat{n}_{j}\left(\hat{a}_{j-1} + \hat{a}_{j+1} \right)+ \text{h.c.} \\
    & - P \sum_{j} \hat{a}_{j}^{\dagger} \hat{a}_{j}^{\dagger} 
         \hat{a}_{j+1} \hat{a}_{j+1} + \text{h.c.},
\end{aligned}
\end{equation}
where $\hat{n}_{j} = \hat{a}_{j}^{\dagger}\hat{a}_{j}$ is the occupation number operator at the lattice site $j$. The first line details the on-site and NN interactions with strengths $U$ and $V$, respectively. The second line proportional to $D$ describes the interaction-induced tunneling process, while the final line corresponds to pair tunneling characterized by magnitude $P$. The interaction
parameters depend  on the overlap integrals between the Wannier functions
\begin{equation}
\label{eq:Gijk}
G_{ijk}=\int\!{\rm d}x\,w_{0}\left(x\right)w_{i}\left(x\right)w_{j}\left(x\right)w_{k}\left(x\right),
\end{equation}
as well as on the scattering strengths $g_{pq}$, $p,q\in\{1,2\}$ between the bare atomic states $|1 \rangle$ and $|2 \rangle$.
By choosing a convenient parametrization for these scattering strengths 
\begin{equation}
\label{eq:gParams}
\begin{pmatrix}g_{11} & g_{12}\\
g_{21} & g_{22}
\end{pmatrix}=\begin{pmatrix}g_{0}+g_{z} & g_{0}+g_{x}\\
g_{0}+g_{x} & g_{0}-g_{z}
\end{pmatrix}\,,
\end{equation}
the effective interaction strengths can be compactly written as
\begin{equation}
\label{eq:UVDP-defs}
\begin{aligned}
    U &= (2g_{0} + g_{x}) G_{000}, \\
    V &= 2g_{0} G_{011}, \\
    D &= -g_{z} G_{001}, \\
    P &= \frac{g_{x}}{2} G_{011}.
\end{aligned}
\end{equation}
 Thus, both the interaction-induced tunneling (described by $D$)  and the pair tunneling ($P$) are due to the state-dependence of the atom-atom interaction.

\section{Many-body ground state analysis}
\label{sec:many-body-analysis}

\begin{figure}[h]
\includegraphics[width=0.4\textwidth]
{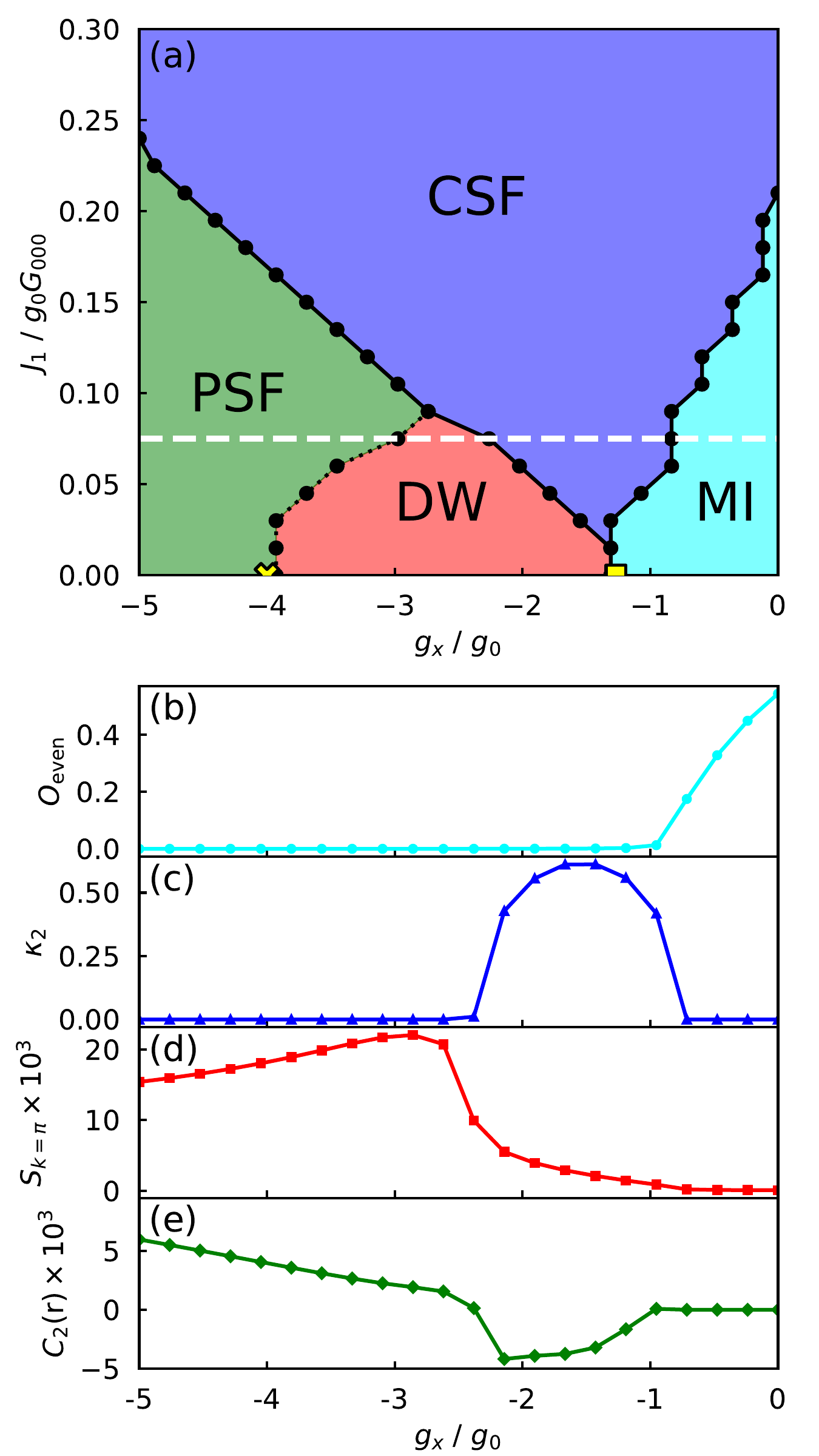}
\caption{(a) Ground state phase diagram for the Hamiltonian given by Eqs.~\eqref{eq:HMain}-\eqref{eq:HIntMain} with $D=0$ (i.e., $g_z=0$), showing the dependence of the phases on parameter ratios $g_x/g_0$ and $J_1/g_0G_{000}$. Solid lines indicate sharp phase transitions, while dotted line specifies a BKT transition. Yellow markers for $J_1=0$ (corresponding to the horizontal $g_x/g_0$ axis) indicate exact transition points determined by the spin mapping, explained in Sec.~\ref{sec:spin-mapping}. The $g_x$ dependence of: (b) even parity order parameter $O_{\rm even}\left(\Delta  j\right)$, defined in Eq.~\eqref{eq:even-parity}; (c) chirality-chirality correlation $\kappa_2 \left(\Delta  j\right)$, defined in Eq.~\eqref{eq:chirality-chirality-corr}; (d) the structure factor $S\left(k\right)$ at $k=\pi$, defined in Eq.~\eqref{eq:struct-fact}; (e) pair correlation function $C_2\left(\Delta  j\right)$, defined in Eq.~\eqref{eq:C2-corr};
for $J_1/g_0G_{000}=0.075$ and $\Delta j = 159$. Note that $C_1\left(\Delta  j\right)$ is non-zero for long ranges in the CSF region, but it is not shown for simplicity.
}
\label{fig:phase-diag}
\end{figure}

In what follows, we perform density matrix renormalization group (DMRG) calculations~\cite{WhitePRL1992} to investigate ground states of the Bose-Hubbard type Hamiltonian, given by Eqs.~\eqref{eq:HMain}-\eqref{eq:HIntMain}. We study a system with $L=200$ lattice sites containing $200$ atoms corresponding to the unit filling and open boundary conditions. A maximum bond dimension of $\chi=500$ was used, which yielded a maximum truncation error of $10^{-6}$. All expectation values, i.e., correlation functions and order parameters, were evaluated in the central $160$ lattice sites to minimize edge effects. Calculations were performed with the ITensors.jl library~\cite{ITensor, ITensor-r0.3}, which is written in the Julia programming language~\cite{Julia-2017}.

We also make a few simplifying assumptions. Firstly, we impose a three-body constraint to stabilize the pair superfluid phase and prevent collapse to a bright soliton. In other words, each lattice site may only have a maximum of 2 bosons. This constraint can be engineered via three-body losses or interactions~\cite{Daley2009PRL, Cirac2010PRA, WesselPRL2011}. We use the Wannier function of the $s$ band by assuming occupation in the lowest energy band, since it is the simplest to realize experimentally. Although the $p$ band physics has been studied and created experimentally~\cite{HemmerichKock_2016, HemmerichWirth2011, Wang2021} for different ultracold atom systems, it would be experimentally challenging for the current setup. Next, we assume $J_2/J_1\approx -3.565$, which corresponds to an experimentally feasible parameter regime considered in previous work~\cite{EdvinasScipost}. Finally, we set the density-dependent tunneling strength at zero ($D=0$), which is the case for equal scattering lengths for atoms in the same internal states ($g_{11}=g_{22}$ or $g_z=0$). A sufficiently strong density-dependent hopping will yield a superfluid, coexisting with a pair superfluid~\cite{EckholtNJP2009}.

\subsection{Quantities of interest}
\label{subsec:quant_of_int}

\begin{table}[H]
\centering
\small
\begin{tabular}{|c|c||c|c|c|c|c|}
\hline
Phase & Acronym 
& $\mathcal{O}_{\mathrm{even}}^{2}$
& $S(k=\pi)$ 
& $C_{1}$ 
& $C_{2}$ 
& $\kappa_2$ \\
\hline
Mott insulator & MI 
& $\neq 0$ & $\approx 0$ & $\approx 0$ & $\approx 0$ & $\approx 0$ \\
\hline
Density wave & DW 
& $\approx 0$ & $\neq 0$ & $\approx 0$ & $\approx 0$ & $\approx 0$ \\
\hline
Superfluid & SF 
& $\approx 0$ & $\approx 0$ & $\neq 0$ & $\neq 0$ & $\approx 0$ \\
\hline
Pair superfluid & PSF 
& $\approx 0$ & $\approx 0$ & $\approx 0$ & $\neq 0$ & $\approx 0$ \\
\hline
Chiral superfluid & CSF 
& $\approx 0$ & $\approx 0$ & $\neq 0$ & $\neq 0$ & $\neq 0$ \\
\hline
\end{tabular}
\begin{centering}
 
\par\end{centering}
\caption{Characterization of different quantum phases with correlation functions
and order parameters~\cite{LucaZakrzewskiReview2025}.
}
\end{table}

Before we discuss the phase diagram presented in Fig.~\ref{fig:phase-diag}, we present various quantities, which we use to characterize the different quantum phases. Due to the Mermin-Wagner theorem~\cite{MerminWagner1966, HohenbergPR1967}, one-dimensional systems exhibit no true condensate density and instead, they display quasicondensation, characterized by algebraically decaying correlation functions~\cite{MuellerPRB2022, Giamarchi1DBook}.

Typical quasicondensation can be characterized by power-law decay of the single particle Green's function~\cite{LucaZakrzewskiReview2025, Huber2013PRB}
\begin{equation}
\label{eq:C1_corr}
C_1\left(\Delta j\right) = \langle \hat{a}_{j}^{\dagger}\hat{a}_{j+\Delta j}\rangle\,.
\end{equation}
Practically, it is usually enough to evaluate the correlation function $C_1(\Delta j)$ at sufficiently large distances $\Delta j$. For insulating phases such as Mott insulators (MI) and density waves (DW), one has $C_1(\Delta j)\approx 0$ due to an exponential decay of the correlation function.
For superfluids (SF) and chiral superfluids (CSF), one expects a finite value of the single particle Green's function, since one has only an algebraic decay.

Additionally, we consider the pair correlation function~\cite{Huber2013PRB, LucaScipost2024, EckholtPRA2008, EckholtNJP2009}
\begin{equation}
\label{eq:C2-corr}
C_{2}\left(\Delta j\right)=\langle \hat{a}_{j}^{\dagger}\hat{a}_{j}^{\dagger}\hat{a}_{j+\Delta j}\hat{a}_{j+\Delta j}\rangle\,.
\end{equation}

A power-law decay of the pair correlation function $C_{2}\left(\Delta j\right)$ with an exponential decay in the single particle correlation function $C_1\left(\Delta j\right)$ indicates the pair superfluid (PSF). Note that the algebraic decay of $C_{2}\left(\Delta j\right)$ alone is not enough to indicate the existence of a PSF, as the SF and CSF phases also exhibit this property. Alternatively, the PSF could also be identified by a finite odd parity~\cite{LucaScipost2024}. However, the odd parity is also non-zero for the DW and thus, cannot determine the PSF-DW transition.

In regimes of large repulsive on-site interactions, the ground state becomes a Mott insulator (MI), which is characterized by the even parity order parameter
\begin{equation}
\label{eq:even-parity}
\mathcal{O}_{\mathrm{even}}^{2} \left(\Delta j\right)=\left\langle \prod_{j_0\le j\le j_0+\Delta j}e^{i\pi\delta\hat{n}_{j}}\right\rangle\,,
\end{equation}
where $j_0$ is the lattice site index of the left side of    the bulk and $\delta\hat{n}_{j} = \hat{n}_j-\bar{n}$ is the boson number deviation.

On the other hand, significant repulsive nearest-neighbor (NN) interactions give rise to spontaneous breaking of the discrete translational symmetry. In particular, we find period-2 density wave (DW) order, which is indicated by a maximum at $k=\pi$ in the structure factor~\cite{LucaZakrzewskiReview2025,Huber2013PRB,Burba2025CommPhys}
\begin{equation}
\label{eq:struct-fact}
S\left(k\right) = \frac{1}{L^2}\sum_{j,l} \langle \hat{n}_j \hat{n}_l\rangle e^{ik(j-l)}\,.
\end{equation}
Note that one could also use a staggered density to incidate the DW order
\begin{equation}
\label{eq:stag-density}
\delta N = \frac{1}{L} \sum_{j} (-1)^{j} \left( \langle \hat{n}_{j} \rangle - \bar{n} \right),
\end{equation}
where $\bar{n} = L^{-1}\sum_j \langle \hat n_j\rangle$ is the average boson filling. We find that $\delta N$ qualitatively agrees with $S\left(k=\pi\right)$ for an odd number of lattice sites, as expected.

Finally, we consider the current-current correlation function~\cite{Li_2016, Burba2025CommPhys, GoldmanPRX2023, DiLiberto2023, Santos2025arXivChiral, Barbiero2025arXivSS}
\begin{equation}
\label{eq:chirality-chirality-corr}
\kappa_2\left(\Delta j\right) = \langle\hat{\kappa}_j\hat{\kappa}_{j+\Delta j}\rangle\,,
\end{equation}
where $\hat{\kappa}_j = i \left(\hat{a}_j\hat{a}^\dagger_{j+1} - \hat{a}^\dagger_j\hat{a}_{j+1} \right)$ is the current operator between sites $j$ and $j+1$. Finite current-current correlations indicate spontaneous breaking of time-reversal symmetry. A long-range non-zero $\kappa_2$ in addition to the power law decay in $C_1$ indicates the existence of a chiral superfluid (CSF).

\subsection{Phase diagram}

The main results are summarized in Fig.~\ref{fig:phase-diag}. The top panel, Fig.~\ref{fig:phase-diag}(a), shows the different ground state quantum phases as a function of Hamiltonian parameters $g_0$, $g_x$ and $J_1$.
The ground state is a Mott insulator for a
sufficiently small $J_1/g_0G_{000}$
and a sufficiently small $g_x/g_0$, which corresponds to strong repulsive on-site interactions, as can be seen in Eq.~\eqref{eq:UVDP-defs}. Increasing $J_1/g_0G_{000}$, one has significant geometric frustration, which yields a chiral superfluid, as expected~\cite{Li_2016, Burba2025CommPhys, GoldmanPRX2023, DiLiberto2023, Santos2025arXivChiral, Barbiero2025arXivSS}. If instead $g_x/g_0$ is decreased, the MI transforms into a DW since the strength of the on-site interaction weakens with respect to the NN interaction. If $g_x/g_0$ is further decreased, the pair tunneling amplitude increases and the pair superfluid emerges. As $g_x/g_0$ decreases, the relative tunneling strength of the MI-CSF transition  $J_1/g_0G_{000}$ decreases since the repulsive on-site interaction, which favors the insulating phase, weakens. Furthermore, the critical tunneling strength corresponding to the PSF-CSF transition increases since the pair hopping becomes stronger.

Finally, as the relative tunneling strength $J_1/g_0G_{000}$ increases, the interaction parameter $g_x/g_0$ of the DW-PSF transition also becomes larger. This can be understood by considering the perturbative (with respect to $J_1$) corrections affecting a simplified DW state  $|2020\ldots\rangle$. The second-order NN tunneling processes ($|2020\ldots\rangle\rightarrow|1120\ldots\rangle\rightarrow|2020\ldots\rangle$) yield a negative correction to the NN interaction term. Thus, including tunneling favors the PSF over the DW by lowering the repulsive NN interactions. Note that the second-order NNN tunneling processes are forbidden for the simplified DW since the intermediate state would have a triply occupied state which is disregarded due to large three-body losses and/or interactions.

The bottom panel, Fig.~\ref{fig:phase-diag}(b), displays different quantities indicating quantum phases, which were discussed in Sec.~\ref{subsec:quant_of_int}, and their dependence on $g_x/g_0$ for fixed $J_1/g_0G_{000}=0.075$. The CSF transitions (MI-CSF and CSF-DW) are sharply defined, which is seen as a sudden change of order parameters. On the other hand, the DW-PSF transition is smoother and less clearly defined. In Sec.~\ref{sec:spin-mapping} we will show that the DW-PSF transition is a Berezinskii-Kosterlitz-Thouless (BKT) quantum phase transition (QPT).

Note that for $g_x/g_0<-2$, the on-site interactions become attractive. Without the three-body restriction, sufficiently attractive on-site interactions would yield a bright soliton state. This has been checked with DMRG calculations with a higher boson dimension. Thus, the three-body restriction (or strongly repulsive three-body interactions) is essential for stabilizing the PSF.

\section{Spin mapping for high barrier regime}
\label{sec:spin-mapping}

Let us now consider our many-body Hamiltonian in the high barrier regime, where the tunneling processes can be neglected, i.e., $H_{\rm{tun}}=0$ or $J_1=J_2=0$. Since we also neglect density-induced hopping, the Hamiltonian simplifies to
\begin{equation}
\label{eq:HInt}
\begin{aligned}
\hat{H}_{\rm hb} &=
	\frac{U}{2} \sum_{j} \hat{n}_{j} (\hat{n}_{j} - 1) 
    + V \sum_{j} \hat{n}_{j} \hat{n}_{j+1} \\
    & - P \sum_{j} \hat{a}_{j}^{\dagger} \hat{a}_{j}^{\dagger} 
         \hat{a}_{j+1} \hat{a}_{j+1} + \text{h.c.}\,,
\end{aligned}
\end{equation}
where $U$, $V$ and $P$ are defined by Eq.~\eqref{eq:UVDP-defs}. 
In the regime where on-site interactions are weakly repulsive or attractive, the low-energy physics can be accurately described by states with zero or two bosons occupying a single site, labeled by $|0\rangle_j$ and $|2\rangle_j$, respectively. 
Since single-particle tunneling processes are negligible, there is no connection to the single occupation sector $|1\rangle_j$ and we can neglect the latter subspace.

Within the subspace in question, it is possible to map our bosonic model onto a spin-$1/2$ chain by
\begin{equation}
|0\rangle_{j} \rightarrow |\downarrow\rangle_{j}\,,\quad |2\rangle_{j} \rightarrow |\uparrow\rangle_{j}\,.
\end{equation}
The bosonic operators are related to the pseudospin-$1/2$ operators through
\begin{equation}
\hat{n}_{j} \rightarrow 2\hat{S}_{j}^{z}+1\,,\quad \hat{a}_{j}^{2} \rightarrow \sqrt{2}\hat{S}_{j}^{-}\,,\quad  \hat{a}_{j}^{\dagger2} \rightarrow \sqrt{2}\hat{S}_{j}^{+}\,,
\end{equation}
and one obtains the  Hamiltonian
\begin{equation}
\label{eq:Ham-spin-full}
\begin{aligned}
\hat{H}_{\rm hb} &=
	U\sum_{j}\hat{S}_{j}^{z}\left(2\hat{S}_{j}^{z}+1\right) \\
    & + V\sum_{j}\left(2\hat{S}_{j}^{z}+1\right)\left(2\hat{S}_{j+1}^{z}+1\right) \\
    & -2P\sum_{j}\left(\hat{S}_{j}^{+}\hat{S}_{j+1}^{-}+\hat{S}_{j+1}^{+}\hat{S}_{j}^{-}\right)\,.
\end{aligned}
\end{equation}

Since we are working at unit boson filling (canonical ensemble), the pseudospin system is characterized by a zero total magnetization as there is an equal number of zero and double occupied lattice sites. As the total magnetization is fixed, we can drop the on-site interaction term proportional to $U$ in Eq.~\eqref{eq:Ham-spin-full}, which corresponds to a global shift in energy. For the same reason, one can neglect the constant and linear (with respect to spin operators) contributions to the NN interaction term proportional to $V$ in Eq.~\eqref{eq:Ham-spin-full}. Consequently, we find that our model maps onto the XXZ Hamiltonian~\cite{YangYang1, YangYang2, YangYang3}
\begin{equation}
\label{eq:XXZ-Ham}
\hat{H}_{{\rm hb}}=-J_{\rm hb}\sum_{j}\left(\hat{S}_{j}^{x}\hat{S}_{j+1}^{x}+\hat{S}_{j}^{y}\hat{S}_{j+1}^{y}+\Delta_{\rm hb} \hat{S}_{j}^{z}\hat{S}_{j+1}^{z}\right)\,,
\end{equation}
where $J_{\rm hb}=4P=2g_{x}G_{011}$ and $\Delta_{\rm hb}=-V/P=-4g_{0}/g_{x}$. Since we consider a regime where $g_0>0$ and $g_x<0$, it follows that $J_{\rm hb}<0$, which corresponds to the antiferromagnetic (AFM) XXZ model.

This mapping allows us to compare DMRG simulations with the well-established XXZ phase diagram. In particular, for the antiferromagnetic case $J_{\rm hb}<0$, the N\'eel phase ($\Delta_{\rm hb} > 1$) corresponds to a DW in the bosonic language, while the XY phase ($|\Delta_{\rm hb}| \le 1$) corresponds to a PSF. The ferromagnetic (FM) phase ($\Delta_{\rm hb} < -1$) lies outside the parameter region where our approximation is valid. Instead, as $g_x/g_0$ is increased, the DW changes into a MI.

From this model the MI-DW transition point for $J_1/g_0G_{000}=0$ can be determined analytically to be $g_x/g_0\approx-1.268$, which agrees well with DMRG numerics, as can be seen from the yellow marker on the horizontal $g_x/g_0$ axis in Fig.~\ref{fig:phase-diag}(a). This was done by comparing AFM Bethe ansatz energy with the MI energy, which is explained in Appendix~\ref{appendix:Bethe-ansatz}.
Finally, we observe that the spin mapping also holds very well in the DW and PSF regions. Details of evaluating the  accuracy of the approximation are given in Appendix~\ref{appendix:subsp-approx}.

\section{Concluding remarks}
\label{sec:discussion}

The results presented in this work demonstrate that the state-dependent subwavelength dark-state lattices provide a versatile route toward engineering strongly correlated bosonic models with qualitatively new capabilities. In contrast to conventional optical lattices, where the correlated hopping terms are negligible, the state-dependent lattices generate significant pair tunneling and density-dependent processes originating from state-dependent interactions and controlled overlap of Wannier functions. Furthermore, the tripod scheme allows for the creation of an unusual lattice, where the amplitude of the NNN tunneling is larger in magnitude and staggered with respect to the NN hopping.

The phase diagram obtained from DMRG simulations reveals an intricate interplay between geometric frustration and pair hopping. For strong on-site repulsion, the system forms a Mott insulator, while large nearest-neighbor repulsion stabilizes the density-wave phase. As the pair hopping strength increases, a pair superfluid phase emerges in which single-particle correlations decay exponentially but pair correlations exhibit quasi-long-range order. Finally, when the staggered tunneling amplitudes become significant, the ground state develops finite current-current correlations, signaling spontaneous breaking of time-reversal symmetry and the onset of a chiral superfluid.

In the high-barrier regime, where the single-particle and density-induced tunnelings are negligible, the low-energy physics can be accurately described by states with zero or two bosons occupying a single site. In this regime, the system admits a mapping onto an antiferromagnetic XXZ spin-1/2 chain. This mapping allows us to identify the DW-PSF transition as the N\'eel-XY transition of the XXZ model, as well as to determine the MI-DW transition point analytically by comparing the Bethe ansatz N\'eel energy with the MI energy. The excellent agreement between these analytical predictions and the DMRG results validates the spin mapping. 
Beyond this regime, one cannot restrict to the aforementioned subspace and the full bosonic model must be studied.

Experimentally, most of the required tools are within reach of current ultracold-atom technology. The subwavelength barriers for dark state atoms of a simpler $\Lambda$ configuration have already been implemented~\cite{Wang2018,Tsui2020}, and the extension to the tripod coupling scheme is feasible. The state-dependent scattering lengths can be tuned via Feshbach resonances~\cite{GrimmRMP2010}. The relevant observables are accessible using established measurement techniques: the density-wave order can be detected via structure factors measured in time-of-flight~\cite{Greiner2002, Lin2009b, Altuntas2023}; pair correlations can be probed through photoassociation measurements~\cite{Zhou2009PRA, EckholtNJP2009} and chiral order can be inferred from the momentum distribution~\cite{Burba2025CommPhys, Santos2025arXivChiral}.

Several extensions of the present work are possible. Allowing finite density-induced tunneling will introduce additional complexity to the competition between geometric frustration and pair hopping. If one relaxes the three-body constraint, there would be formation of clusters and solitonic states. It is also possible to consider $p$-band Wannier states, where the pair hopping would be even stronger. Additionally, one could couple bands together via Floquet engineering~\cite{Eckardt17RMP} and study the multiband model~\cite{Luhmann_2012, FleischhauerPRA2011, HansPeterPRL2010}. Finally, one could study dynamics of doublons and holes, which has been done in other ultracold atom systems~\cite{kim2025multiparticlequantumwalksdipoleconserving, Leonard2023}.

In summary, we have proposed and analyzed a realistic scheme that unifies geometric frustration and strong pair hopping within a single experimentally accessible platform. The resulting competition between chiral order and pair condensation enriches the landscape of quantum phases beyond the standard Bose-Hubbard model~\cite{LucaZakrzewskiReview2025, Greiner2002, MonienPRB1998, EjimaEPL2011, EjimaPRA2012} and establishes subwavelength dark-state lattices as a powerful tool for quantum simulation of exotic many-body physics.

\section*{Acknowledgements}

D.B. thanks Luca Barbiero and Nitya Cuzzuol for helpful discussions.
This project was supported by the Research Council
of Lithuania (RCL) Grant No. S-LJB-24-2 and the JSPS Bi-lateral Program No. JPJSBP120244202. D.B. utilized the computing resources of the Scientific Computing and Data Analysis section of Core Facilities at Okinawa Institute of Science and Technology Graduate University (OIST), as well as the resources at the High Performance Computing Center (HPCC), ``HPC Sauletekis'' in Vilnius University, Faculty of Physics.

\appendix

\section{Derivation of interaction Hamiltonian}
\label{appendix:HIntDerivation}

\subsection{Interaction Hamiltonian}

Let us consider the interaction term $\hat{H}_{\rm int}$ featured in the Hubbard Hamiltonian (\ref{eq:HMain}). Since the amplitude of the Rabi frequency $\Omega_3(x)$ is much larger than that of the remaining two Rabi frequencies $\Omega_1(x)$ and $\Omega_2(x)$, the dark state atoms populate mostly the internal ground states $|1\rangle$ and $|2\rangle$. Therefore, the interaction Hamiltonian can be written in terms of bosonic field operators for these two components
\begin{equation}
    \hat{H}_{\rm int} = \sum_{p,q\in\{1,2\}} g_{pq} \int \!{\rm d}x\, \pd{p}(x)\pd{q}(x)\pc{q}(x)\pc{p}(x).
\end{equation}
Using the following parametrization for the atom-atom interaction strength

\begin{equation}
\label{eq:gParams'}
\begin{pmatrix}g_{11} & g_{12}\\
g_{21} & g_{22}
\end{pmatrix}=\begin{pmatrix}g_{0}+g_{z} & g_{0}+g_{x}\\
g_{0}+g_{x} & g_{0}-g_{z}
\end{pmatrix}\,,
\end{equation}
the interaction Hamiltonian becomes
\begin{equation}
\hat{H}_{\mathrm{int}}=\hat{H}_{\mathrm{int}}^{\left(0\right)}+\hat{H}_{\mathrm{int}}^{\left(z\right)}+\hat{H}_{\mathrm{int}}^{\left(x\right)}\,,\label{eq:H_int-separation}
\end{equation}
with
\begin{align}
\hat{H}_{\mathrm{int}}^{\left(0\right)} & = g_{0}\sum_{p,q\in\{1,2\}}\int \!{\rm d}x\, \pd{p}(x)\pd{q}(x)\pc{q}(x)\pc{p}(x), \\
\hat{H}_{\mathrm{int}}^{\left(z\right)} & = g_{z}\sum_{\substack{p\in\{1,2\}}
}\left(-1\right)^{p+1}\!\int \!{\rm d}x\, \pd{p}(x)\pd{p}(x)\pc{p}(x)\pc{p}(x), \\
\hat{H}_{\mathrm{int}}^{\left(x\right)} & = 2g_{x}\int \!{\rm d}x\, \pd{1}(x)\pd{2}(x)\pc{2}(x)\pc{1}(x).
\end{align}

In terms of the field operators corresponding to the symmetric or antisymmetric superpositions of the atomic ground states
\begin{equation}
    \pc{\pm}(x)=\frac{1}{\sqrt{2}}\left(\pc{1}(x)\pm\pc{2}(x)\right),
\end{equation}
the terms comprising the interaction Hamiltonian read
\begin{align}
\hat{H}_{\mathrm{int}}^{\left(0\right)}&=g_{0}\sum_{p,q\in\{+,-\}}\int\!dx\ \pd{p}(x)\pd{q}(x)\pc{q}(x)\pc{p}(x), \\
\hat{H}_{\mathrm{int}}^{\left(z\right)}&=g_{z}\!\int\!dx\,\left[\pds{+}(x)+\pds{-}(x)\right]\pc{+}(x)\pc{-}(x)+ \text{h.c.}, \\
\hat{H}_{\mathrm{int}}^{\left(x\right)}&=\frac{g_{x}}{2}\!\int\!dx\,\left[\pds{+}(x)-\pds{-}(x)\right]\left[\pcs{+}(x)-\pcs{-}(x)\right].
\end{align}
Expanding the operators $\pc{\pm}(x)$ in the basis of the localized Wannier functions via Eq.~(\ref{eq:WannierBasis}), the interaction terms take the form:
\begin{align}
\hat{H}_{\mathrm{int}}^{\left(0\right)} &=g_{0}\sum_{\alpha,\beta\in\left\{ 0,1\right\} }\sum_{ijkl} G^{2i+\alpha,2j+\beta}_{2k+\beta,2l+\alpha} \ \hat{a}_{2i+\alpha}^{\dagger}\hat{a}_{2j+\beta}^{\dagger}\hat{a}_{2k+\beta}\hat{a}_{2l+\alpha}, \\
\hat{H}_{\mathrm{int}}^{\left(z \right)} &= g_{z}\sum_{ijkl} \Big( G^{2i+1,2j+1}_{2k+1,2l} \ \hat{a}_{2i+1}^{\dagger}\hat{a}_{2j+1}^{\dagger}\hat{a}_{2k+1}\hat{a}_{2l}\\
 & + G^{2i,2j}_{2k+1,2l} \ \hat{a}_{2i}^{\dagger}\hat{a}_{2j}^{\dagger}\hat{a}_{2k+1}\hat{a}_{2l} \Big)+ \text{h.c.}, \\
 \hat{H}_{\mathrm{int}}^{\left(x\right)} &=  
 \frac{g_{x}}{2}\sum_{ijkl} \Big( G^{2i+1,2j+1}_{2k+1,2l+1} \ \hat{a}_{2i+1}^{\dagger}\hat{a}_{2j+1}^{\dagger}\hat{a}_{2k+1}\hat{a}_{2l+1} \\
 & -\Big[G^{2i+1,2j+1}_{2k,2l} \ \hat{a}_{2i+1}^{\dagger}\hat{a}_{2j+1}^{\dagger}\hat{a}_{2k}\hat{a}_{2l} + \text{h.c.} \Big] \\
 & +G^{2i,2j}_{2k,2l} \ \hat{a}_{2i}^{\dagger}\hat{a}_{2j}^{\dagger}\hat{a}_{2k}\hat{a}_{2l} \Big).
\end{align}
Here the sums run over $i,j,k,l \in \mathbb{Z}$ and the Wannier function integral is defined by 
\begin{equation}
G^{ij}_{kl} \equiv \int\mathrm{d}x\,w_{i}^*\left(x\right)w_{j}^*\left(x\right)w_{k}\left(x\right)w_{l}\left(x\right),
\end{equation}
with superscripts $i$ and $j$ referring to the complex conjugated Wannier functions. 

We now use the basic symmetry properties of the Wannier functions and the lattice to simplify the notation and reduce the number of independent integrals.
First, for a one-dimensional system with a symmetric potential the maximally localized Wannier functions are real \cite{Kohn1959Aug,Marzari2012Oct}, so the complex conjugation in the integral can be dropped. We indicate this by lowering the indices, $G_{ijkl}$. Second, the system is invariant under translations by the lattice constant $a$, which is equivalent to shifting all Wannier function indices by two, i.e.~$G_{ijkl} = G_{i+2,j+2,k+2,l+2}$. This invariance allows one index to be eliminated by choosing an arbitrary reference point in the lattice. To reduce notational clutter, we set this reference at index 0 and define $G_{ijk} \equiv G_{0ijk}$ as in Eq.~(\ref{eq:Gijk}) of the main text. Finally, $G_{ijk}$ is symmetric under index permutations, hence we order the indices in monotonically increasing order.

As only the neighboring Wannier functions have a significant overlap, the Hamiltonian expression can be simplified. In particular, for symmetric Wannier functions the relevant set of independent overlap integrals reduces to $G_{000}$, $G_{001}$ and $G_{011}$. The interaction terms then become
\begin{align*}
\hat{H}_{\mathrm{int}}^{\left(0\right)} &=g_{0}G_{000}\sum_{j}  \hat{n}_{j}(\hat{n}_{j}-1) + g_{0} G_{011}\sum_{j} \hat{n}_{j}\hat{n}_{j+1}, \\
\hat{H}_{\mathrm{int}}^{\left(z \right)} &= g_{z} G_{001}\sum_{j}  \hat{a}_{j}^{\dagger} \hat{n}_{j}\left(\hat{a}_{j-1} + \hat{a}_{j+1} \right)+ \text{h.c.}, \\
 \hat{H}_{\mathrm{int}}^{\left(x\right)} &=  
 \frac{g_{x}}{2}G_{000} \sum_{j} \hat{n}_{j}(\hat{n}_{j}-1) \\
 & -\frac{g_{x}}{2}G_{011}\sum_{j} \hat{a}_{j+1}^{\dagger}\hat{a}_{j+1}^{\dagger}\hat{a}_{j}\hat{a}_{j} + \text{h.c.}.
\end{align*}
 Summing up the terms we arrive at the interaction Hamiltonian given by Eq.~(\ref{eq:HIntMain}) in the main text.

\subsection{Overlap integrals}
Let us estimate the overlap integrals $G_{ijk}$ between the Wannier functions  featured in the interaction Hamiltion. As the probability of tunneling through the subwavelength barriers is much less than unity, one can approximate the Wannier function $w_{j}\left(x\right)$ centered at $x= ja/2$ in terms of the wave-function of a particle confined by infinitely high potential barriers at $x= (j\pm 1)a/2$.
\subsubsection{$s$ band}

For the $s$-band considered in the present work, the approximate Wannier functions illustrated in Fig.~\ref{fig:FigBrickWall}(b) can be written as
\begin{equation}
w_{j}^{(s)}\left(x\right)\approx\left(-1\right)^{j+1}\sqrt{\frac{2}{a}}\sin\left(\frac{\pi}{a}\left(x-ja/2\right)\right)\,.
\end{equation}
This leads to the overlap integrals
\begin{equation}
G_{000}^{(s)}\approx\int_{0}^{a}{\rm d}x\,\left(w_{0}^{(s)}\left(x\right)\right)^{4}=\frac{3}{2a}\,,
  \end{equation}

\begin{equation}
G_{001}^{(s)}\approx\int_{a/2}^{a}{\rm d}x\,\left(w_{0}^{(s)}\left(x\right)\right)^{3}w_{1}^{(s)}\left(x\right)=-\frac{1}{\pi a}\,,
\end{equation}

\begin{equation}
G_{011}^{(s)}\approx\int_{a/2}^{a}{\rm d}x\,\left(w_{0}^{(s)}\left(x\right)\right)^{2}\left(w_{1}^{(s)}\left(x\right)\right)^{2}=\frac{1}{4a}\,,
\end{equation}
with their ratios given by
\begin{equation}
G_{001}^{(s)}/G_{000}^{(s)}\approx-\frac{2}{3\pi}\approx-0.2122\,,
\end{equation}

\begin{equation}
G_{011}^{(s)}/G_{000}^{(s)}\approx\frac{1}{6}\approx0.1667\,,
\end{equation}

\begin{equation}
G_{001}^{(s)}/G_{011}^{(s)}\approx-\frac{4}{\pi}\approx-1.2732\,.
\end{equation}

\subsubsection{Arbitrary band}

For generality, one can study the Wannier functions for the $\alpha$-th ($\alpha\in \mathbb{N}$) Bloch band given by
\begin{equation}
w_{j}^{(\alpha)}\left(x\right)\approx\left(-1\right)^{j+1}\sqrt{\frac{2}{a}}\sin\left(\frac{\alpha\pi}{a}\left(x-ja/2\right)\right)\,,
\end{equation}
for which one will find that $G_{000}$ does not depend on $\alpha$
\begin{equation}
G_{000}^{(\alpha)} = \frac{3}{2a}\,.
\end{equation}
On the other hand, $G_{011}^{(\alpha)}/G_{000}^{(\alpha)}$ can be found to alternate between $1/6$ and $1/2$ as $\alpha$ is increased. The expression for $G_{011}^{(\alpha)}$ is
\begin{equation}
G_{011}^{(\alpha)}=\frac{2+\left(-1\right)^{\alpha}}{4a}\,,    
\end{equation}
and $G_{001}$ is given by
\begin{equation}
G_{001}^{(\alpha)} =
\begin{cases}
\displaystyle
(-1)^{\frac{\alpha+1}{2}}\frac{1}{\alpha \pi a},
& \text{if } \alpha \text{ is odd}, \\[8pt]
\displaystyle
(-1)^{\frac{\alpha}{2}+1}\frac{3}{4a},
& \text{if } \alpha \text{ is even}.
\end{cases}
\end{equation}

Thus, to achieve large values of $G_{011}$ and $G_{001}$ (which is required for accessing the exotic regimes), while keeping experimental feasibility in mind, it is best to restrict attention to the $s$ or $p$ bands, since higher bands are more difficult to realize. In terms of the pair hopping strength, the $p$ band is the most favorable.

\section{MI-DW transition point via the Bethe ansatz energy}
\label{appendix:Bethe-ansatz}

In this section we find the exact transition point between the DW and MI for $J_1=0$. To achieve this, note that the DW in the bosonic picture corresponds to  the N\'eel phase in the pseudospin picture. Thus, one can calculate the N\'eel phase energy via the Bethe ansatz solution for the AFM XXZ~\cite{YangYang1, YangYang2, YangYang3} Hamiltonian ($J_{\rm hb}<0$,
$\Delta_{\rm hb}>1$). The aforementioned energy per lattice site is given by~\cite{Takahashi_1999}
\begin{equation}
e_{{\rm AFM}}=-\frac{J_{\rm hb}\Delta_{\rm hb}}{4}+J_{\rm hb}\sinh\phi\left[\frac{1}{2}+2\sum_{n=1}^{\infty}\frac{1}{e^{2n\phi}+1}\right]\,,
\end{equation}
where $\phi=\cosh^{-1}\left(\Delta_{\rm hb}\right)$.

Since $J_{\rm hb}=4P=2g_{x}G_{011}$ and $\Delta_{\rm hb}=-V/P=-4g_{0}/g_{x}$ (see Sec.~\ref{sec:spin-mapping}), one
obtains
\begin{equation}
e_{{\rm AFM}}=2g_{0}G_{011}\left(1+\frac{g_{x}}{g_{0}}\sinh\phi\left[\frac{1}{2}+2\sum_{n=1}^{\infty}\frac{1}{e^{2n\phi}+1}\right]\right)\,,
\end{equation}
where $\phi=\cosh^{-1}\left(-4\left(g_{x}/g_{0}\right)^{-1}\right)$.

\begin{figure}[tb]
\centering \includegraphics[width=0.75\columnwidth]{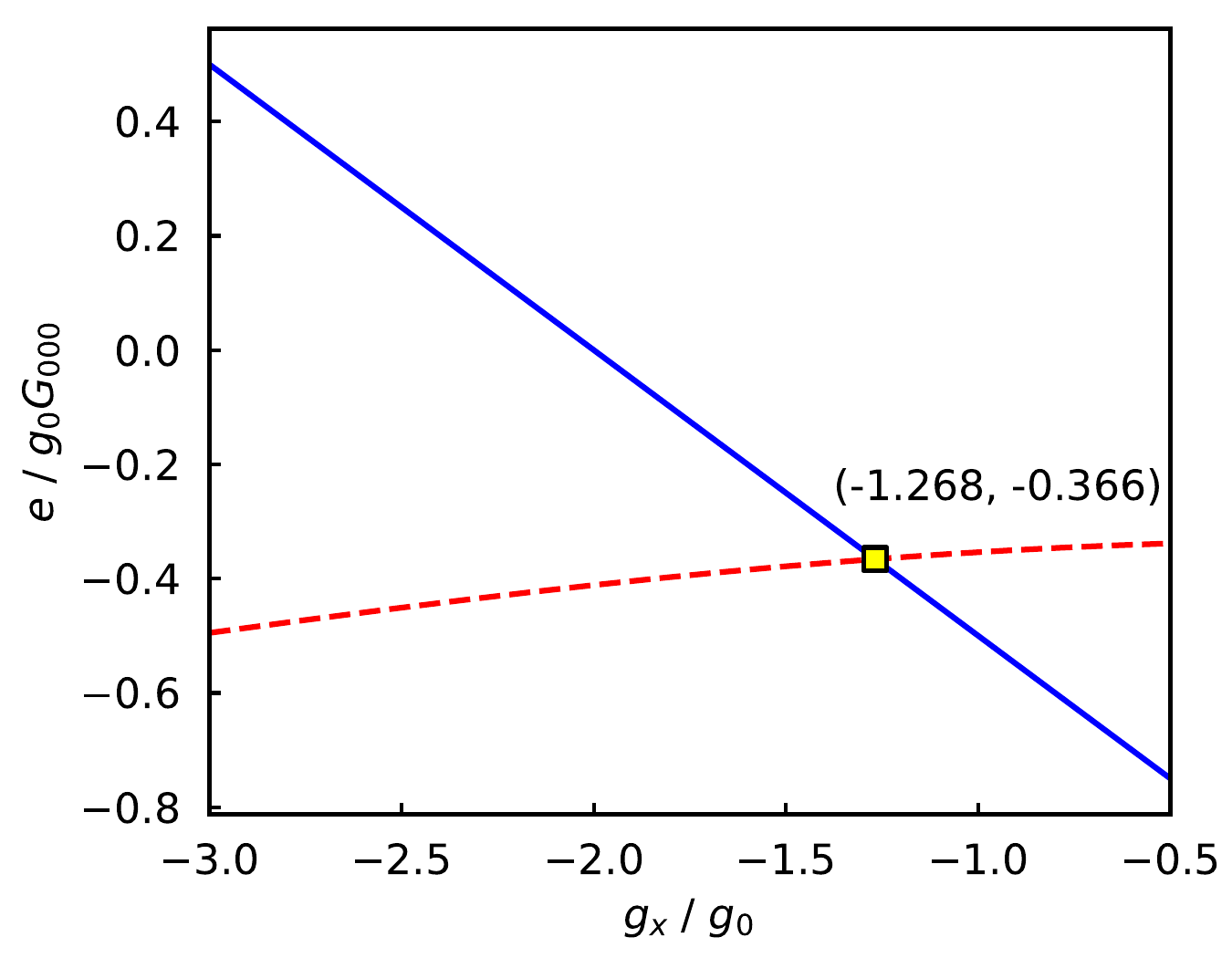}
\caption{Energies per site $e$ vs. the interaction strength ratio $g_{x}/g_{0}$.
The Mott insulator energy per site $e_{{\rm MI}}$ is represented by a
solid blue line, while AFM energy per site $e_{{\rm AFM}}$ is represented
by a dashed red line.}
\label{fig:XXZ_phase_curve_schematic-1}
\end{figure}

The Mott insulator $|1111\ldots\rangle$ energy per site would simply be $\tilde{e}_{{\rm MI}}=V$.
However, we also need to take into account the global energy shift,
which we omitted deriving the spin Hamiltonian in Sec.~\ref{sec:spin-mapping}
\begin{equation}
\Delta H=U\sum_{j}\hat{S}_{j}^{z}\left(2\hat{S}_{j}^{z}+1\right)+V\sum_{j}\left(4\hat{S}_{j}^{z}+1\right)\,.
\end{equation}
Since we are working at zero magnetization ($\langle \hat{S}_{j}^{z}\rangle=0$),
the terms proportional to $\hat{S}_{j}^{z}$ do not contribute to $\Delta H$. Using $\left(\hat{S}_{j}^{z}\right)^{2}=1/4$,
one obtains a constant energy shift per site $\Delta e=U/2+V$. Thus,
to make a comparison between the energies, we compute the shifted
Mott insulator energy per site as
\begin{equation}
e_{{\rm MI}}=\tilde{e}_{{\rm MI}}-\Delta e=-U/2\,.
\end{equation}
Using $U=\left(2g_{0}+g_{x}\right)G_{000}$, one obtains 
\begin{equation}
e_{{\rm MI}}=-g_{0}G_{000}\left(1+\frac{1}{2}\frac{g_{x}}{g_{0}}\right)\,,
\end{equation}
and in this way, the transition point is given by
\begin{equation}
e_{{\rm MI}}=e_{{\rm AFM}}\,.
\end{equation}
Solving numerically, one finds that the transition point is at $g_{x}/g_{0}\approx-1.26829$
with $e_{{\rm MI}}=e_{{\rm AFM}}\approx-0.36586$, which can be seen in Fig.~\ref{fig:XXZ_phase_curve_schematic-1}. This agrees very well with the
DMRG numerics, as can be seen from the yellow marker in Fig.~\ref{fig:phase-diag}.

\section{Accuracy of subspace approximation}
\label{appendix:subsp-approx}

The accuracy of projection in the manifold of single or double occupancies can be quantified by the subspace probability $P_{\bar Q}=\langle\hat{\bar Q}\rangle$, with $ \hat{\bar Q}=\prod_j \hat{\bar Q}_j$,
where $\hat{\bar Q}_j=|0\rangle_j\langle0|_j + |2\rangle_j\langle2|_j$ is the lattice site projection operator onto the space of occupation numbers 0 and 2.
Yet, the numerically obtained ground states have a small, but non-negligible number of singly occupied lattice sites. Thus, we introduce a modified projection operator $\hat{Q}=\prod_j \hat{Q}_j$, where $\hat{Q}_j$ is characterized by the softening parameter $\alpha_Q \ll L$
\begin{equation}
\hat{Q}_j=|0\rangle_j\langle0|_j + |2\rangle_j\langle2|_j + e^{-\alpha_Q/L} |1\rangle_j\langle 1|_j\,,
\end{equation}
with $L$ being a number of lattice sites. Smaller values of $\alpha_Q/L$ allow for more singly occupied lattice sites.
Again, the accuracy of the approximation is quantified by the subspace probability $P_Q = \langle \hat{Q}\rangle$. 

\begin{figure}[tb]
\includegraphics[width=0.4\textwidth]
{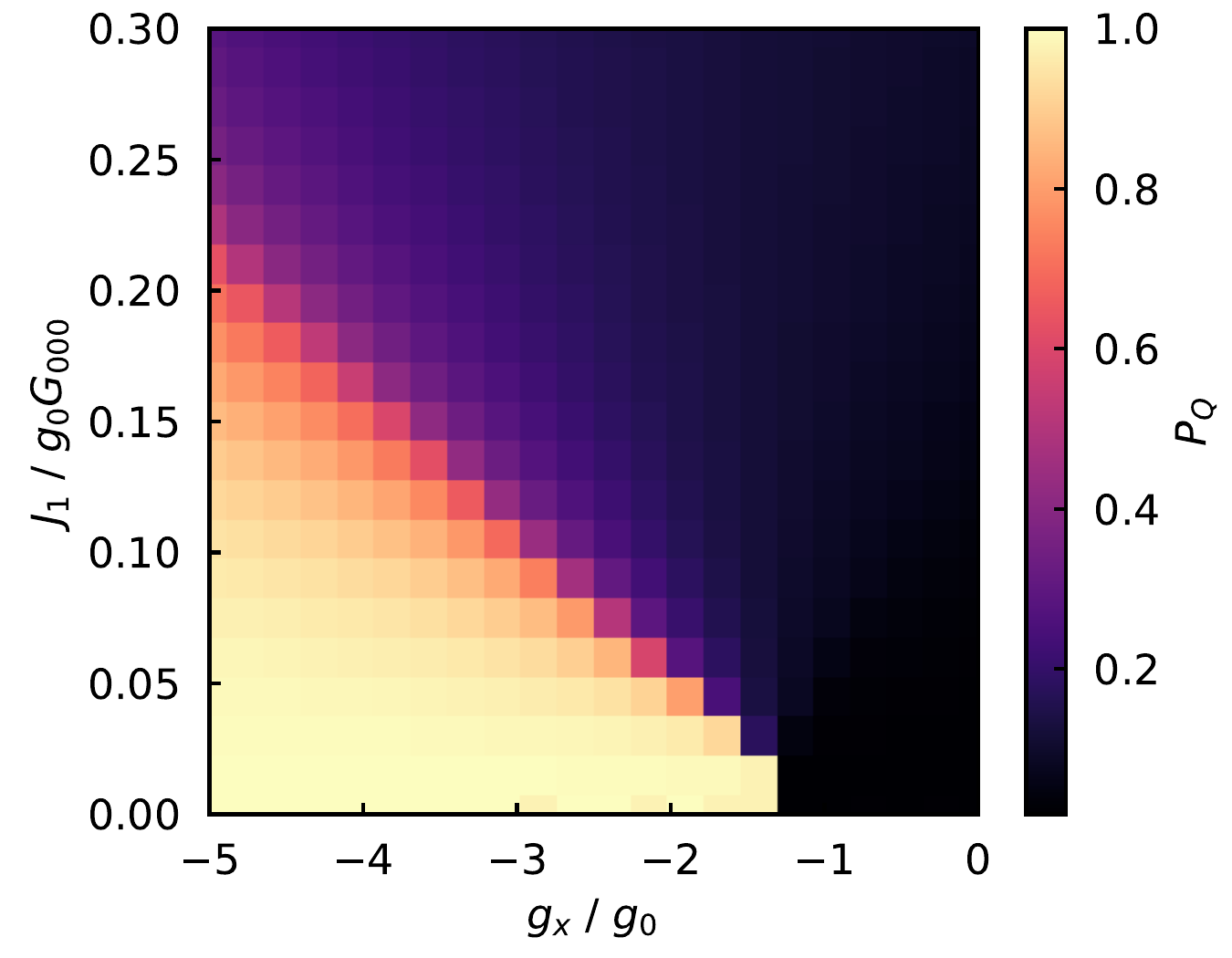}
\caption{Dependence of the subspace probability $P_Q$ on the parameter ratios $g_x/g_0$ and $J_1/g_0G_{000}$ for the softening parameter $\alpha_Q=5.0$. The DMRG calculations were performed for $200$ particles in a lattice with $L=200$ sites and open boundary conditions. Local boson dimension is set to 3 (maximum of 2 bosons per site). A maximum bond dimension $\chi=500$ was used in the DMRG calculations.
}
\label{fig:subsp-approx}
\end{figure}
In Fig.~\ref{fig:subsp-approx}, we show the dependence of the subspace probability $P_Q$ on the Hamiltonian parameters $g_0$, $g_x$ and $J_1$. We observe that the accuracy of the projection holds very well ($P_Q\approx 1$) in the DW and PSF regions. As the tunneling amplitude ratio $J_1/g_0G_{000}$ increases, the approximation becomes less accurate, which is signified by $P_Q <1$.

\bibliography{main}

\end{document}